\documentclass[runningheads]{llncs}
\usepackage{makecell}
\usepackage[T1]{fontenc}
\usepackage{multirow}
\usepackage{marvosym}
\usepackage{hyperref}
\usepackage[perpage,stable]{footmisc}
\renewcommand{\thefootnote}{}
\newcommand{\ieno}{\textit{i}.\textit{e}.}

\newcommand{\egno}{\textit{e}.\textit{g}.}

\usepackage{graphicx}
\usepackage{xcolor}  
\usepackage{soul}
\usepackage{amsmath}
\usepackage{amssymb}

\begin{document}
\title{RTN: Reinforced Transformer Network for Coronary CT Angiography Vessel-level Image Quality Assessment}
%
%
\titlerunning{RTN}

\author{Yiting Lu\inst{1}\footnotemark[1]
\and
Jun Fu\inst{1}\footnotemark[1]
\and
Xin Li\inst{1}
\and 
Wei Zhou\inst{1}
\and 
Sen Liu\inst{1}
\and 
Xinxin Zhang\inst{2}
\and 
Congfu Jia\inst{2}
\and  Ying Liu\inst{2}
\and Zhibo Chen\inst{1}\footnotemark[2]
} 
\authorrunning{Y. Lu et al.}
\institute{ University of Science and Technology of China, Hefei, Anhui, China\\
\email{\{luyt31415, fujun, lixin666, weichou\}@mail.ustc.edu.cn}\\
\email{elsen@iat.ustc.edu.cn}\\
 \email{chenzhibo@ustc.edu.cn}
\and 
The First Affiliated Hospital of Dalian Medical University, Dalian, Liaoning, China}
%
%
\maketitle              
\renewcommand{\thefootnote}{\fnsymbol{footnote}}
\footnotetext[1]{The first two authors contribute equally to this work.}
\footnotetext[2]{Corresponding Author}
\begin{abstract}

Coronary CT Angiography (CCTA) is susceptible to various distortions (e.g., artifacts and noise), which severely compromise the exact diagnosis of cardiovascular diseases. The appropriate CCTA Vessel-level Image Quality Assessment (CCTA VIQA) algorithm can be used to reduce the risk of error diagnosis. 
The primary challenges of CCTA VIQA are that the local part of coronary that determines final quality is hard to locate. 
To tackle the challenge, we formulate CCTA VIQA as a multiple-instance learning (MIL) problem, and exploit \textbf{T}ransformer-based  \textbf{MIL} backbone (termed as T-MIL) to aggregate the multiple instances along the coronary centerline into the final quality.
However, not all instances are informative for final quality. There are some quality-irrelevant/negative instances intervening the exact quality assessment(\egno, instances covering only background or the coronary in instances is not identifiable). Therefore, we propose a \textbf{P}rogressive \textbf{R}einforcement learning based \textbf{I}nstance \textbf{D}iscarding module (termed as PRID) to progressively remove quality-irrelevant/negative instances for CCTA VIQA. 
Based on the above two modules, we propose a \textbf{R}einforced \textbf{T}ransformer \textbf{N}etwork (RTN) for automatic CCTA VIQA based on end-to-end optimization.
Extensive experimental results demonstrate that our proposed method achieves the state-of-the-art performance on the real-world CCTA dataset, exceeding previous MIL methods by a large margin.
\end{abstract}
%
%
\section{Introduction}
Coronary Computed Tomography Angiography (CCTA) technique plays an indispensable role in the diagnosis of cardiovascular diseases for providing vital visual clues.
However, the CCTA images are easily degraded by various factors (\ieno, patient breathing motion artifacts and insufficient contrast agent dose) and contain hybrid distortions~\cite{li2020learning,liu2020lira}, which inevitably affects the subsequent analysis of expert doctors~\cite{ghekiere2017imageCCTASurvey}. For example, when artifacts appear in the coronary artery stenosis, it is difficult for doctors to distinguish whether the vessel is stenosis or not~\cite{leipsic2010adaptive_itereco}. To ensure accurate diagnosis, it is necessary to provide doctors with high-quality CCTA images. Therefore, there is an urgent need to develop CCTA Vessel-level Image Quality Assessment (CCTA VIQA) algorithms that can be used to automatically quantify the perceptual image quality of CCTA.


With the rapid development of machine learning, the seminal work~\cite{nakanishi2018automated} maps hand-crafted global and local features (\ieno, noise, contrast, misregistration scores, and un-interpretability index) of coronary artery onto image quality scores through machine learning algorithms. However, its input features are not rich since they only include four types of image characteristics, which always causes the sub-optimal performance and lacks of enough flexibility.
Also, quality metric~\cite{liu2022swiniqa,liu2021liqa} designed for natural image are not suitable for medical image.
During the dataset annotation process, the professional doctors only provide the vessel-level label when browsing the complete CT. So no position labels are provided for quality relevant regions and the key local parts that determine the vessel-level quality are hard to locate, which shows CCTA VIQA is an obvious weakly-supervised problem~\cite{zhou2018weaksurvey}. So the quality relationship between various local parts of coronary artery in CCTA image can be excavated by modeling CCTA VIQA as a MIL problem.  Therefore, we propose Transformer-based MIL backbone (T-MIL) in CCTA VIQA. Specifically, since the quality of CCTA images is only associated with the coronary artery, we utilize the centerline tracking algorithm~\cite{wolterink2019CNNTracker} to detect the regions of coronary artery. Then we define 3D cubes cropped along the coronary centerline as instances. Finally, the discriminative features from multiple instances extracted by 3D convolutional neural networks are aggregated into the quality space through the latest network architecture, \ieno, transformer. Recently, there are various MIL aggregators in MIL methods, like attention~\cite{ilse2018attentionMIL,CLAM,li2021dsmil}, RNN~\cite{campanella2019MIL_RNN}, sparse convolution~\cite{lerousseau2021sparseconvmil}, and graph~\cite{graph}. Specially, transformer-based MIL frameworks~\cite{lee2019settransformer,myronenko2021accountingtransformer,shao2021transMIL,yu2021mil-vt} have achieved remarkable success in a broad of medical tasks, such as whole slide image classification.

Although the instances (\ieno, cubes) have covered all possible quality-associated contents, the quality-irrelevant contents also 
infiltrate the instances severely, which is detrimental for the estimation of overall quality. For instance, the quality-related cubes only take a small proportion of all cubes. According to our observation, there are three typical cases of quality-irrelevant instances \ieno, the instance that does not match the vessel-level label, the coronary in instances is not identifiable, and the instance contains only background. To remove these negative instances while mining the most informative instances, we propose a \textbf{P}rogressive \textbf{R}einforcement learning based \textbf{I}nstance \textbf{D}iscarding module (termed as PRID) to preserve informative instances as the inputs of the transformer. 
The reinforcement learning (RL) agent from PRID accepts the output feature embedding of transformer as states, and selects one instance to discard. Then we input the new instance set into T-MIL to obtain the states (both in training and testing) for the next iteration and the reward (just for training) to refine current action.
We call the T-MIL together with PRID as \textbf{R}einforced \textbf{T}ransformer \textbf{N}etwork, which is denoted as RTN.

We summarize our contributions as follows. 

\begin{itemize}
    \item To our knowledge, we propose the first fully automatic CCTA VIQA algorithm RTN based on end to end optimization. We formulate the CCTA VIQA as the typical MIL problem, and introduce transformer to aggregate multiple instances and map them to final quality. 
    \item To elide the intervention from quality-irrelevant/negative isntances, we propose a progressive reinforced learning based instance discarding strategy (\ieno, PRID) to  mine the most informative instances for transformer network. 
    \item Extensive experimental results reveal that our proposed RTN achieves the SOTA performance on hospital-built CCTA dataset, exceeding previous MIL methods by a large margin. 
\end{itemize}


\begin{figure}
    \centering
    \includegraphics[width=0.95\linewidth]{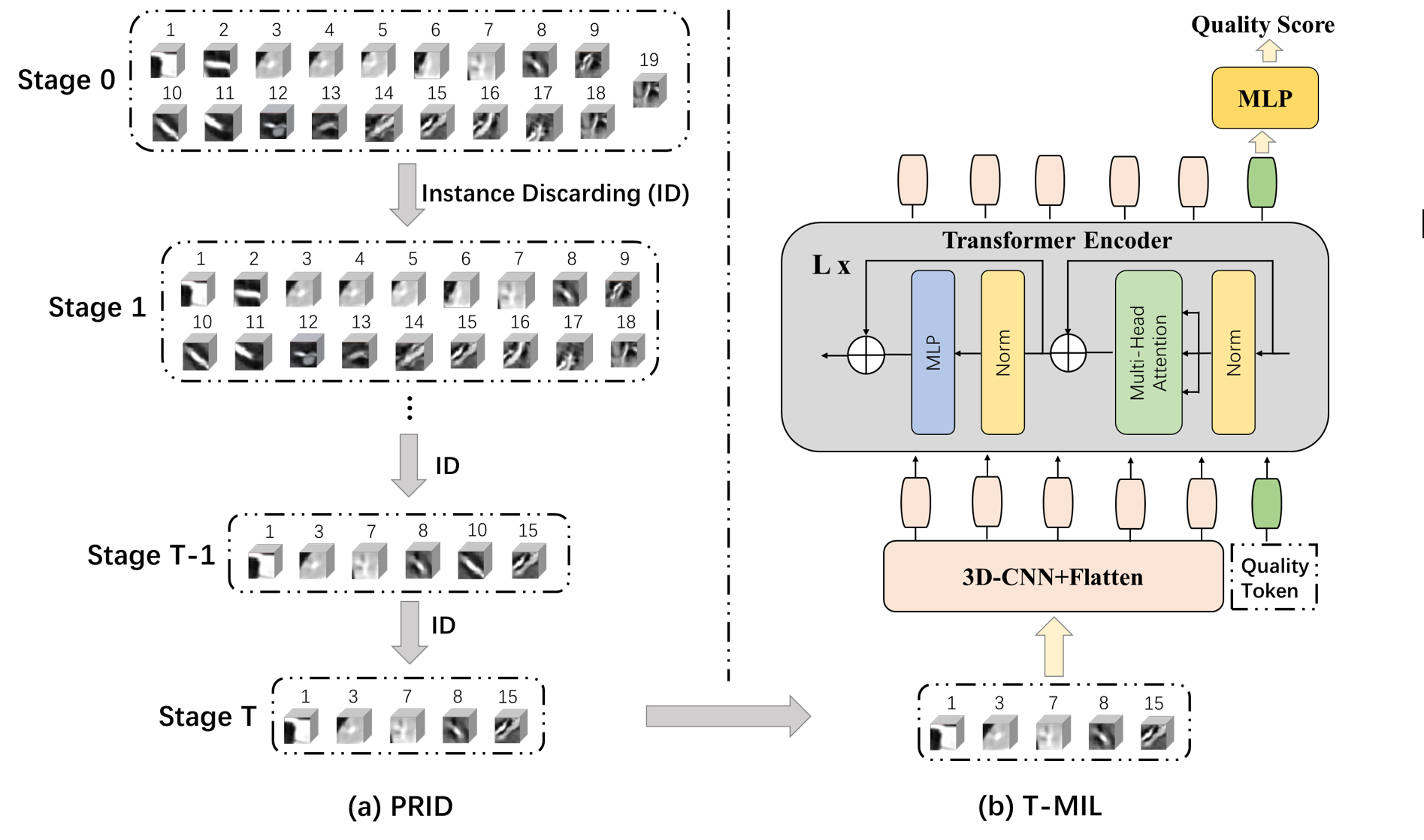}
    \caption{The architecture of RTN, consisting of two basic components: (a) Progressive Reinforcement Learning based Instance Discarding (PRID) and (b) Transformer-based MIL Backbone (T-MIL).}
    \label{fig:framework}
   
\end{figure}

\section{Methods}

Fig.~\ref{fig:framework} depicts the overall framework of RTN for the CCTA VIQA task, which is composed of two basic components \ieno, Progressive Reinforcement Learning based Instance Discarding (PRID) and Transformer-based MIL backbone (T-MIL). 
Given one CCTA image, we first collect the cubes cropped along the coronary centerline as instances. Then PRID module employs a reinforcement learning (RL) agent to determine which instance should be discarded progressively. After obtaining the most informative instances, T-MIL is devoted to classifying the final quality grade in vessel-level. In the following sections, we will clarify the T-MIL and PRID of our RTN from both implementation and principal perspectives.

\subsection{Transformer-based MIL}

MIL is a strong tool to solve weakly-supervised problem. In the definition of MIL, a set of multiple instances can be regarded as a bag and only bag-level label is provided. In the situation of CCTA VIQA, 
the vessel-level quality is only determined by partial local regions of coronary arteries. However, the positions of these local regions cannot be located, which is a weakly-supervised question and consistent with MIL setting. Therefore, in our method, 
we define the $i^{th}$ 3D cube sampled on the centerline of the coronary artery as an instance $x_i$, and the whole coronary artery region is taken as a bag $B=\{x_i|1\le i\le n\}$. Then the perceptual quality $y$ of whole coronary $B$ can be inferred with Eq.~\ref{equ:bag}.
\begin{equation}
y(B) = h(f(x_1),f(x_2),...,f(x_i),...,f(x_n)), ~ 1 \le i \le n.
\label{equ:bag}
\end{equation}
Where,  $x_i \in \mathbb{R}^{C_1\times D \times H \times W}$ is the $i^{th}$ instance in the same bag $B$. T-MIL contains ${f(.)}$ and $h(.)$, which are separately as instance feature extractor and transformer-based aggregator. 
In this paper, the instance feature extractor $f$ is composed of several 3D convolution based residual blocks~\cite{chen2019med3d} and flatten operation.

\subsubsection{Transformer-based Aggregator.}
To capture the long-range dependency between different instances, we employ the transformer architecture in ViT~\cite{dosovitskiy2020vit} as the aggregator of MIL. As shown in Figure~\ref{fig:framework},
each transformer encoder layer is consist of multi-head self-attention (MHSA) layer and feed-forward (FF) layer. We follows the ViT~\cite{dosovitskiy2020vit} and add the quality  token $c_0(B)$ to the instance token groups.
The input token embeddings can be written as: 
\begin{equation}
    Z_{0}=[c_0(B),f(x_1),f(x_2),...,f(x_i),...,f(x_n)], ~ 1 \le i \le n.
\end{equation}

 In MHSA, we firstly transform instance embedding to key $K$, query $Q$ and value $V$, and then calculate the similarity of key and query as attention weight matrix. The matrix's each item means dependencies between any pair of instances. The output of MHSA contains aggregation information, especially quality token embedding that aggregates the contribution of each instance to final vessel-level quality prediction.The full process of the $l^{th}$ transformer layer is as follows, in which LN is layernorm and MLP includes two fully-connected layers with a GELU non-linearity:
\begin{equation}
  \begin{split}
          Z_{l}^{'} &=MHSA(LN(Z_{l-1})), \quad l=1,2,...L\\
          Z_{l}  &=MLP(LN(Z_{l}^{'}))+Z_{l}^{'}, \quad l=1,2...L
  \end{split}
\end{equation}
After feeding input token embedding into $L$ transformer layers, we can obtain the output token embedding $Z_{L}\in\mathbb{R}^{(n+1)\times D}$, in which D is the dimension of the token embedding. The first quality token embedding $c_L(B)=Z_L[0]$ is used to quality classification and following instance embedding $Y_L=Z_L[1:n]$ can be used as the states of PRID. 
\begin{figure}
    \centering
    \includegraphics[width=0.9\linewidth]{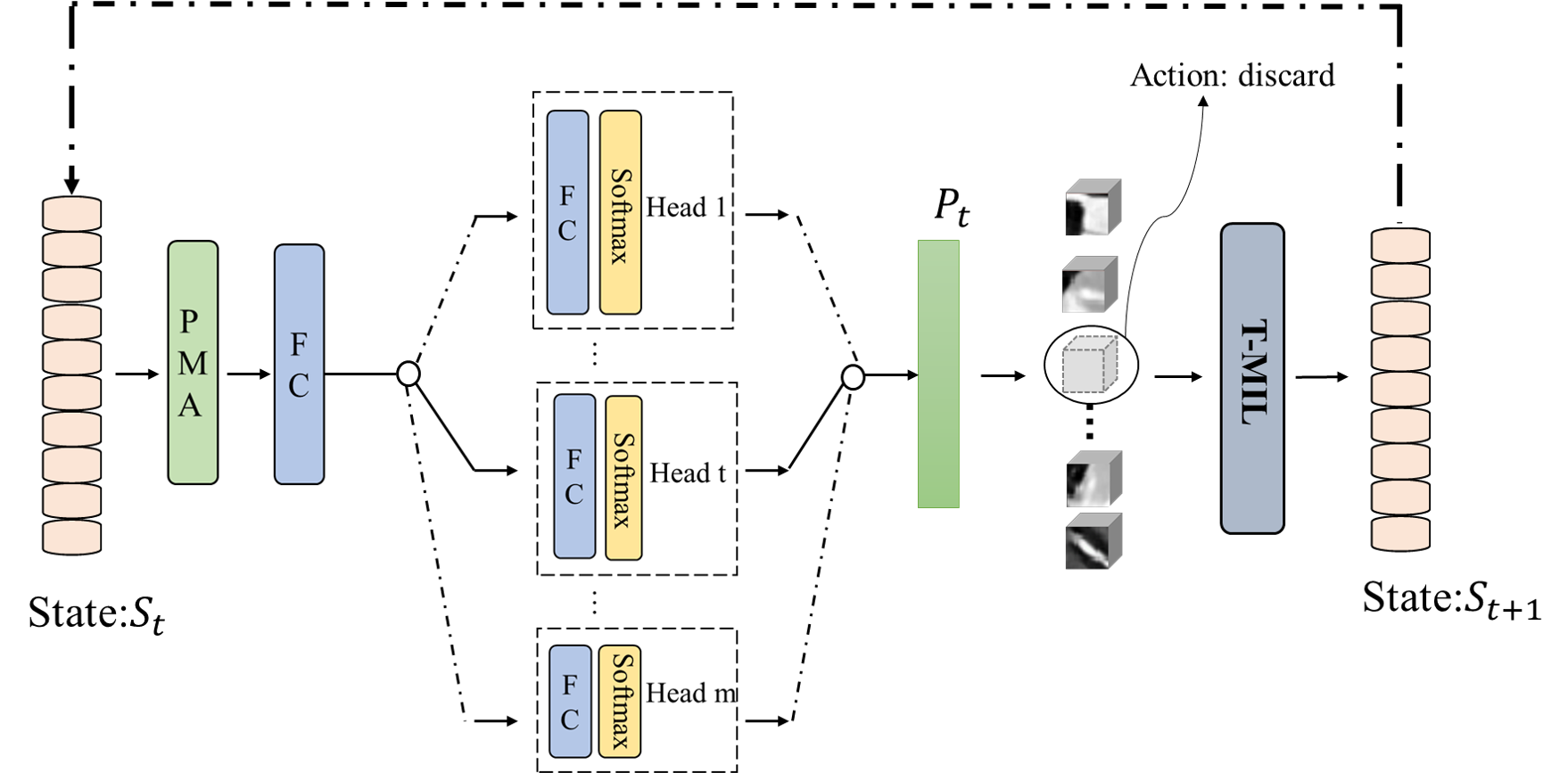}
    \caption{The agent network of PRID, contains common Pooling by Multi-Head Attention(PMA) module and various MLP layers.}
    \label{fig:agent}
    
\end{figure}
\subsection{PRID}

To reduce the intervention of negative instances (\egno, the instance that do not match vessel-level labels or the instance contains only background), we propose to utilize reinforcement learning (RL) agent to adaptively identify them and discard them progressively~\cite{tang2018deep}.
Specifically, we model the process of progressively instance discarding  as a Markov Decision Process (MDP)~\cite{bellman1957markovian,littman2015reinforcement} and introduce a RL agent to obtain the optimal solution for it. The state, action, reward and agent in RL are clarified clearly as follows.


\subsubsection{States.}
As shown in Fig.~\ref{fig:agent}, in the $t^{th}$ iteration, the state $S_t$ is defined as the output instance embedding  $Y_L(t-1)\in\mathbb{R}^{(n-t+1)\times D}$ of the $(t-1)^{th}$ iteration's transformer layer, since  the features captured by transformer are more representative for quality prediction.

\subsubsection{Action.}
Action is the instance index that is discarded within the scope of the instance set. In the $t$-th iteration, the action search space $A=\{1,2,..k,...,n-t+1\}$ is the current instances' index list. The agent's output probability vector $P_t\in\mathbb{R}^{ n-t+1}$can be regarded as the selected distribution of current instance set. Thus we can encode action as multinomial distribution sampling when training and top one sampling when testing, the selected $k$ is equal to action: $k =sample(P_t)$. The state $S$ transforms to $S^{'}$ through the action because of changes in the instance set: $\{x_i\}^{n-t+1} _{i=1} \longrightarrow \{x_i\}^{n-t+1}_{i=1,i\neq k} $.

\subsubsection{Reward.}
The reward $R$ need to reflect the effect of transforming from the state $S$ to $S^{'}$ due to the action. According to the action,  we feed a new instance set into the pre-trained T-MIL and compare the new prediction result with the label to calculate the reward $(t>1)$:
\begin{equation}
   \begin{split}
    R_t &= \begin{cases}
         2, &\quad if~ y_t =label\quad and\quad y_{t-1} =label\\
         1, &\quad if~ y_t =label\quad and\quad y_{t-1} \neq label\\
         -1,&\quad if~ y_t\neq label\quad and\quad y_{t-1} =label\\
         -2,&\quad if~ y_t\neq label\quad and\quad y_{t-1} \neq label
            \end{cases}
   \end{split}
   \label{equ:reward}
\end{equation}
In the first selection, the predict result $y_1$ need to compare with label. If the prediction is correct, give a positive reward ($+1$), otherwise give a negative reward ($-1$). In the next choice, as the Eq.~\ref{equ:reward} shown,  the value of reward is not only related to the accuracy of the current selection's prediction result, but also to the last selection's result. This is because in the MDP problem, the current selection (iteration) is related to the last selection (iteration). 
\subsubsection{Agent.}
As shown in Fig.~\ref{fig:agent}, the agent in PRID receives the states from T-MIL. We first aggregate the $n$ tokens of states into one token through PMA module $PMA(.)$~\cite{lee2019settransformer}. In the $t^{th}$ iteration, this module sets a learnable embedding $I\in\mathbb{R}^{1\times D}$ as a query, and directly regards the instance embedding $Y_{L}(t-1)\in\mathbb{R}^{(n-t+1)\times D}$ as key and value to calculate a attention matrix of $1\times (n-t+1)$ dimension to gather these feature embedding. Similarly, the cross attention here is also implemented in the form of multi-head.
Then we feed the fused token into the MLP head $g_t(.) $ to obtain the probability vector $P_t\in\mathbb{R}^{n-t+1}$. Note that in $t^{th}$ iteration, we will use $t^{th}$ MLP head $g_t(.)$:
\begin{equation}
          P_t=g_t(PMA(Y_{L}(t-1))) 
\end{equation}

\subsubsection{Instance Discarding Strategy.}
The implementation of Instance Discarding Strategy requires above two modules: PRID and T-MIL. In the first stage, we need to pre-train the T-MIL by randomly selecting $n-m$ instances from $n$ instances on the centerline. In the second stage, fix the parameters of T-MIL and update the agent's parameter through $m$ progressive selections through interaction with T-MIL. At each iteration, we can obtain the selected 
index ($k$) probability from distribution $P_t$ and reward $R_t$, so the training loss is 
\begin{equation}
   \begin{split}
   loss= -\sum\nolimits_{t=1}^m log(P_t[k])\times R_t
   \end{split}
\end{equation}
\begin{table}

\centering
\caption{Performance comparisons with state-of-the-arts on the CCTA dataset.}\label{tab1}
\setlength{\tabcolsep}{2mm}{
\begin{tabular}{c|c|c}
\hline MIL methods &  Accuracy & AUC\\
\hline
AttentionMIL~\cite{ilse2018attentionMIL}&  0.7574 & 0.7576 \\
MIL-RNN~\cite{campanella2019MIL_RNN}    &  0.7322 & 0.6842 \\
CLAM~\cite{CLAM}       &  0.7761 & 0.7161\\
DSMIL~\cite{li2021dsmil}       &  0.6917 & 0.5378\\
T-MIL (ours) &  {\bfseries 0.8036} & {\bfseries 0.7658}\\
RTN(PRID+T-MIL) (ours) &  {\bfseries 0.8546} & {\bfseries 0.8461}\\
\hline
\end{tabular}}
\label{table:SOTA}
\end{table}
\section{Experiment}
\subsection{Implementation Details}

Our CCTA VIQA dataset is collected with the help of a partner hospital, where the vessel-level quality labels of each CCTA image are provided by experienced imaging doctors. There are two quality levels in our dataset \ieno, ``1” and ``0”. ``1” means the CCTA image is high-quality and accepted by doctors, while ``0” represents the CCTA image is low-quality and cannot be used for diagnosis. The CCTA VIQA datasets consist of 80 CCTA scans from 40 patients in both systole and diastole, which can be divided into 210 coronary branches by the centerline tracking algorithm ~\cite{wolterink2019CNNTracker}. Therefore, our datasets contain 210 pairs of coronary branches and its corresponding vessel-level quality labels, where the ratio of label ``1” and label ``0” is 114/96. And we plan to make this CCTA VIQA dataset public later.

We adopt the numbers of instances (\ieno, cubes) $n$ in MIL as 19, which are uniformly cropped along the  centerline of coronary branch. All cubes are with the size of $20\times20\times20$ and cover whole coronary branch.
Considering the dataset is small, we also augment the data by moving the cube's center point randomly to three voxels in any direction along 6 neighborhoods as in~\cite{ma2021transformerstenosis}. 
We follow the 5-fold cross validation setting with $80\%$ of data for training and $20\%$ for testing in each split.  
Both T-MIL and PRID are implemented with Pytorch and trained on one NIVDIA 1080Ti GPU. In the training process, we first train T-MIL for 200 epochs with the batchsize 2. Then, we optimize the PRID module for 400 epochs with batchsize 2. 
\begin{table}[]

\centering
\caption{Performance comparison with different discarding numbers and discarding strategies in RTN on the CCTA dataset.}
\label{tab:tab2}
\setlength{\tabcolsep}{3mm}{
\begin{tabular}{c|c|c}
\hline
Discarding Number & PRID(Accuracy/AUC) & Random(Accuracy/AUC) \\ \hline
4                               & 0.7964/0.7777      & 0.7682/0.7409        \\ \hline
9                               & 0.8253/0.7674      & 0.8007/0.7567        \\ \hline
14                              & 0.8546/0.8461      & 0.8107/0.7994        \\ \hline
\end{tabular}}
\label{tab:comparewithrandom}
\end{table}
We utilize two metrics of quality classification at vessel level to measure the effectiveness of the proposed framework: Accuracy and Area Under the Curve (AUC) scores. Moreover, the discarding number of instances $m$ pick 14 as baseline. At the same time, we set 9 and 4 for ablation experiments. 

\subsection{Comparisons with State-of-the-arts}
We compare our methods with the state-of-the-art MIL methods on the CCTA VIQA datasets, including attention-based MIL~\cite{ilse2018attentionMIL}, RNN-based MIL~\cite{campanella2019MIL_RNN}, attention-based and cluster-based MIL~\cite{CLAM}, non-local attention based MIL~\cite{li2021dsmil}. In order to ensure fairness, the feature extraction process of the above methods shares the same two layers of 3D residential blocks. As shown in Table~\ref{table:SOTA}, transformer-based MIL exceeds the second best method CLAM~\cite{CLAM} by 2.75\%, thanks to its better long-range relationship modeling capability.
Furthermore, our proposed RTN achieves the best performance, outperforming previous MIL-based methods by 7.85\%, which reveals the effectiveness of our PRID. In other words, discarding quality-irrelevant instances is vital for CCTA VIQA. See supplementary material, the visualization of index distribution of discarded instance and remained instance shows that 
only limited instances will play a role in CCTA VIQA.
\begin{figure}
    
    \centering
    \includegraphics[width=\linewidth]{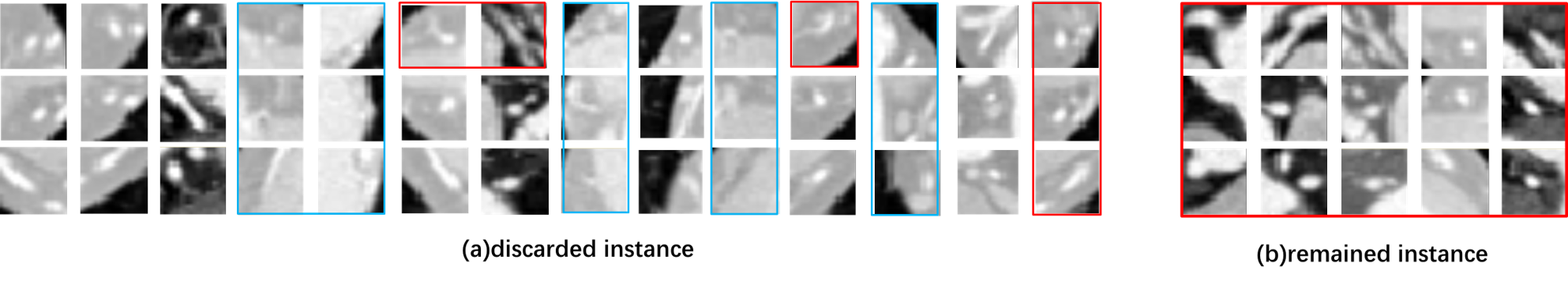}
    \caption{Example of instance discarding with label ``1”. Three rows represent views in axial, sagittal, and coronal orientations from all 3D cubes on the coronary artery, among them, the blue box is the case where the distortion is too serious to judge the coronary by the network, and the red box is the case with obvious distortion.}
    \label{fig:sample}
   
\end{figure}

\subsection{Ablation Study}
In this section, we verify the effectiveness of our proposed PRID from four aspects: the number of discarding instances, discarding strategy, pooling operations and cube size. Table~\ref{tab:tab2} shows the comparison results of different discarding numbers and different discarding strategies. According the results, the discarding number $m=14$ is the best solution. This is because after iterative discarding, the five instances with the most information are retained at last, which will make it easier for network to classify, as shown in Fig.~\ref{fig:sample}. 
We also compare the PRID with random discarding strategy in Table ~\ref{tab:comparewithrandom}. Our PRID exceeds random discarding strategy by a large margin regardless of the discarding number, which reveals the effectiveness of our PRID on instance selection. In Table~\ref{tab3}, we compare the different pooling operations for RL agent in PRID. We can draw a conclusion that PMA has a stronger aggregation ability to input instance embedding. This also shows that it is more explanatory to aggregate tokens through cross attention~\cite{lee2019settransformer}. The comparison of different cube size in Table~\ref{tab3} shows that the cubes with small size cannot cover the whole vessel and the cubes with larger size will contain a little more quality-unrelated content. 

\begin{table}
\centering
\caption{Performance comparison with different pooling module in agent network of RTN on the CCTA dataset and different cube size on one vessel.}\label{tab3}
\setlength{\tabcolsep}{2mm}{
\begin{tabular}{c|c|c|c|c|c}
\hline Pooling Module &  Accuracy & AUC& Crop Size & Accuracy & AUC\\
\hline
PMA &  {\bfseries 0.8546} & {\bfseries 0.8461}& 15 & {0.8042}&{ 0.7459}  \\

Avg Pooling  &  {0.8443} & {0.8257}& 20 & {\bfseries 0.8546} &{ 0.8461}  \\
Max Pooling &  {0.8273} & {0.8198 }& 30 & {0.8510}&{\bfseries 0.8668 }  \\
\hline
\end{tabular}}
\end{table}
\section{Conclusion}
In this paper, we present a novel Reinforced Transformer Network(RTN) model for CCTA VIQA, which contains two modules: Transformer-based MIL backbone (T-MIL) and 
Progressive Reinforcement learning based Instance Discarding module (PRID). T-MIL can solve the challenge that local part of coronary that determines final quality is hard to locate. Moreover, PRID can overcome the intervention from  quality-irrelevant/negative instances. Compared with previous MIL methods, our RTN has achieved great improvement.
In the future, we plan to adaptively select the number of discarded instances, which will continue to be improved in the follow-up work and put into clinical use.

\section{Appendix}
\begin{figure}
    \centering
    \includegraphics[width=1\linewidth]{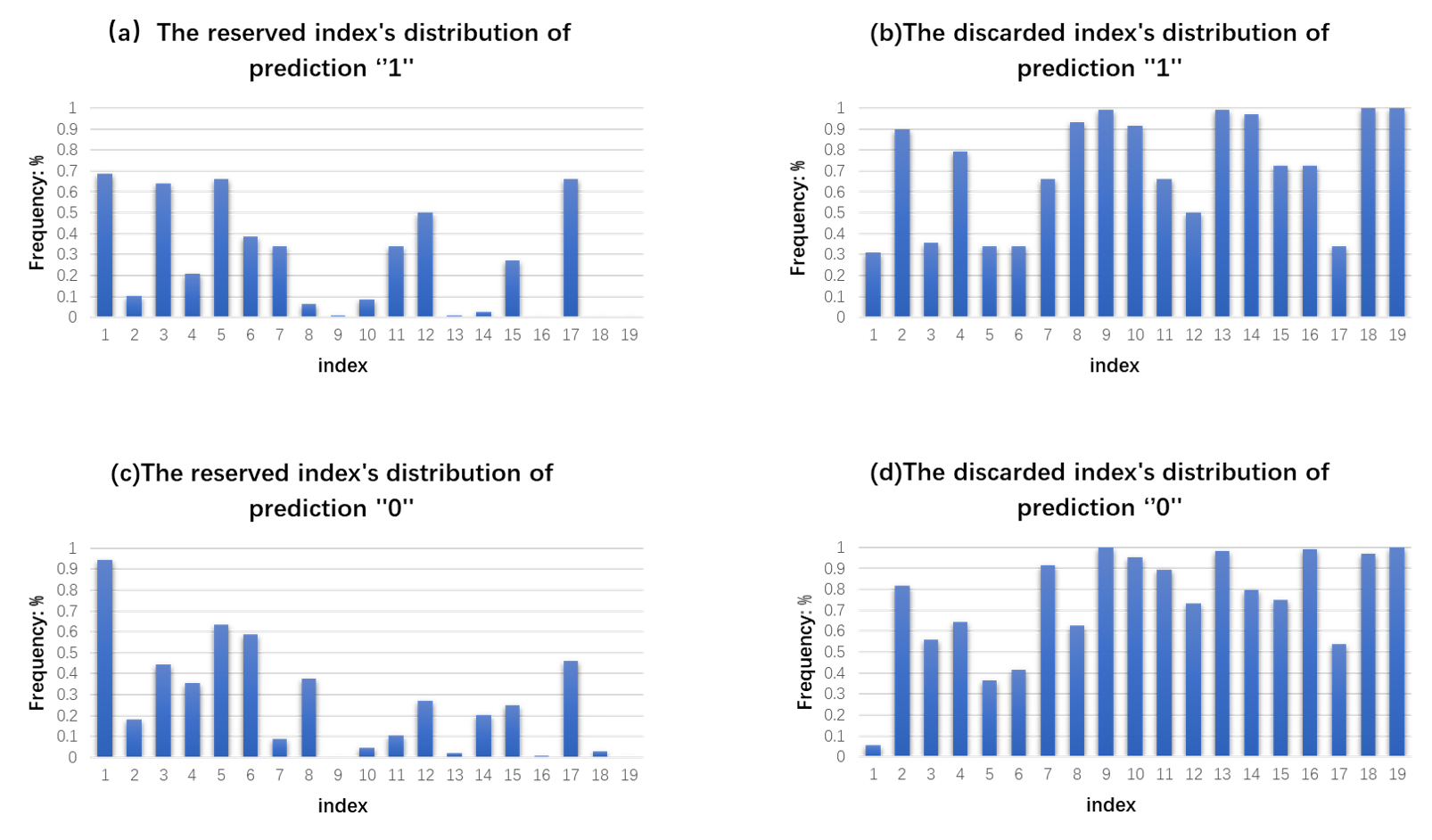}
    \caption{Reserved index's distribution and discarded index's distribution of different prediction result.}

    \label{fig:index_dis}
   
\end{figure}
\begin{figure}
    \centering
    \includegraphics[width=0.9\linewidth]{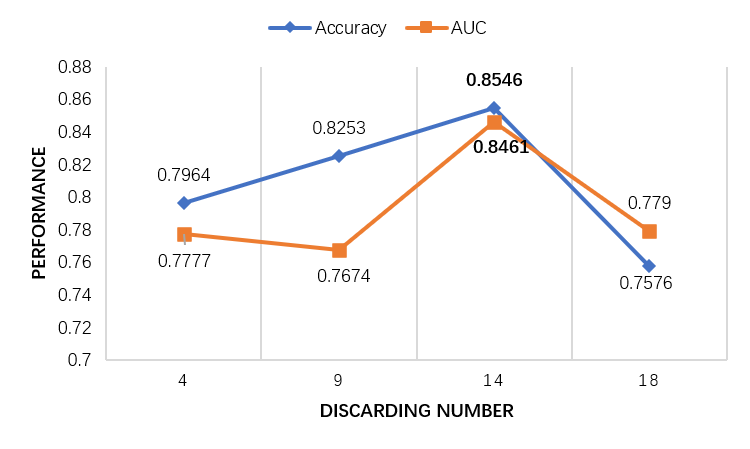}
    \caption{Performance of PRID with different discarding number.}

    \label{fig:index_dis}
   
\end{figure}

\begin{figure}
    \centering
    \includegraphics[width=1\linewidth]{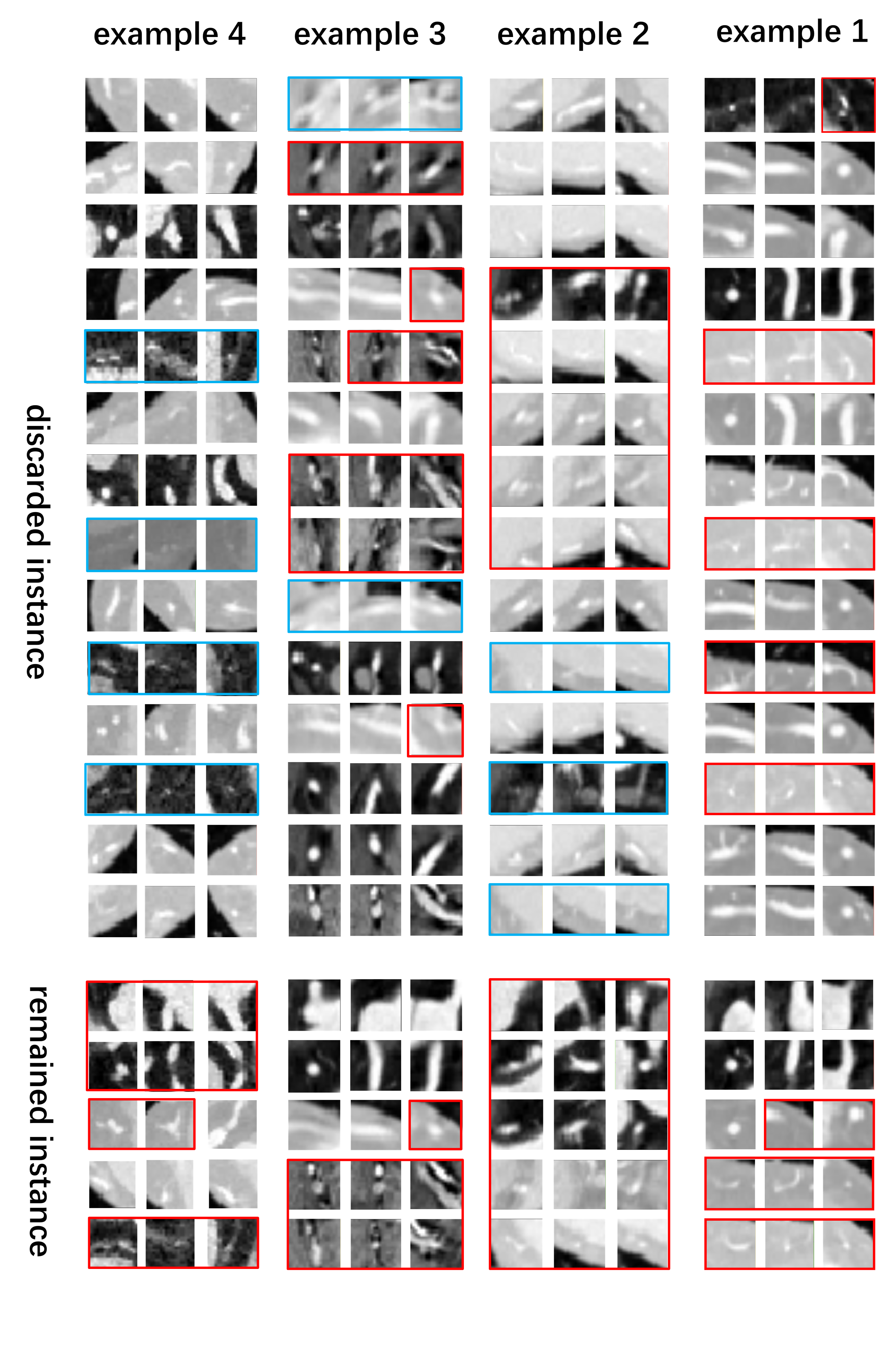}
    \caption{Four examples of instance discarding with label ``1”. Among them, the blue box is the case where the distortion is too serious to judge the coronary by the network or the case that only have background, and the red box is the case with obvious distortion. And instances without boxes are of high quality.}
  
    \label{fig:sample_other}
  
\end{figure}

\clearpage
\subsubsection*{Acknowledgement.} This work was supported in part by NSFC under Grant U1908209, 62021001 and the National Key Research and Development Program of China 2018AAA0101400.

%
%
%

\bibliographystyle{splncs04}
\bibliography{mybibliography}
%

\end{document}